\documentclass[sigconf]{acmart}
\usepackage[acronym]{glossaries}
\usepackage{amsmath,amsfonts, amsthm}
\usepackage{graphicx}
\usepackage{textcomp}
\usepackage{colortbl}
\usepackage{float}
\usepackage{subcaption}
\usepackage{multirow}
\usepackage{multicol}
\usepackage{xspace}
\usepackage{comment}
\usepackage{soul}
\usepackage{stfloats} 
\usepackage{longtable,booktabs}
\usepackage[version=4]{mhchem}
\usepackage{paralist}
\usepackage[compact]{titlesec}
\usepackage{setspace}
\usepackage{xcolor}
\usepackage{hyperref}
\usepackage{annotate-equations}
\usepackage{mathrsfs}
\usepackage{algorithm}
\usepackage{array}
\def\secformat{\bfseries}
\usepackage{listings}

\usepackage[most]{tcolorbox}
\newtcolorbox[auto counter]{finding}[1][]{%
    colback=blue!5,           
    colframe=blue!40,         
    boxrule=0pt,              
    leftrule=2mm,             
    sharp corners,            
    before upper={\textbf{Finding~\thetcbcounter:}~}, 
    fontupper=\normalfont,    
}

\definecolor{neongreen}{rgb}{0.0, 1.0, 0.0} 
\definecolor{neonpink}{rgb}{1.0, 0.07, 0.58} 
\sethlcolor{neonpink}

\usepackage{enumitem}
\definecolor{secblue}{HTML}{00008B}
\definecolor{subsecblue}{HTML}{0000FF}
\definecolor{subsubsecblue}{HTML}{0047AB}
\definecolor{captionblue}{HTML}{0000FF}
\definecolor{tableheader}{HTML}{203864}
\definecolor{paragraph}{HTML}{3B5E7F}
\definecolor{pgtext}{HTML}{0000CD}
\definecolor{hdrgray}{HTML}{BFCDDB}%
\definecolor{ltgray}{HTML}{DCDCDC}%
\definecolor{cellcolor}{HTML}{BFCDDB}

\titleformat{\subsection}{\secformat\color{secblue}}{\thesubsection}{0.5em}{}
\titleformat{\subsubsection}{\secformat\color{subsubsecblue}}{\thesubsubsection}{0.5em}{}

\setlist[itemize]{noitemsep, topsep=0pt, partopsep=0pt,left=0pt}

\titlespacing{\section}{0em}{0.0em}{0em}
\titlespacing{\subsection}{0em}{0.0em}{0em}
\titlespacing{\subsubsection}{0em}{0em}{0em}
\glsdisablehyper
\makeglossaries   
\newacronym{tl}{TL}{Transfer Learning}
\newacronym{hpc}{HPC}{High Performance Computing}
\newacronym{ml}{ML}{Machine Learning}
\newacronym{nlp}{NLP}{Natural Language Processing}
\newacronym{sipt}{M\textsubscript{$\mathcal{R}$}}{\texttt{Stacked Input Alignment Model}}
\newacronym{ipt}{M\textsubscript{$\mathcal{A}$}}{\texttt{Input Alignment Model}}

\newacronym{cn}{$\mathcal{M}odel\mathcal{X}$}{\texttt{Cross Prediction Model}}
\newacronym{sb}{SB}{Intel Sandy Bridge}
\newacronym{bgq}{BGQ}{IBM BG/Q}

\newacronym{mse}{MSE}{Mean Square Error}
\newacronym{nn}{NN}{Neural Network}
\newacronym{lp}{LP}{Linear Probing}
\newacronym{ft}{FT}{Fine-tuning}
\newacronym{ind}{IND}{In-Distribution}
\newacronym{ood}{OOD}{Out-of-Distribution}
\newacronym{fsl}{FSL}{Few-Shot Learning}
\newacronym{sssp}{SSSP}{Single-Source Shortest Path}
\newacronym{llms}{LLMs}{Large Language Models}
\newacronym{llm}{LLM}{Large Language Model}
\usepackage{tabularray}
\definecolor{sg}{HTML}{45B39D}

\definecolor{sgl}{HTML}{A2D9CE}
\definecolor{bl}{HTML}{5499C7}
\definecolor{bll}{HTML}{A9CCE3}
\definecolor{or}{HTML}{D35400}
\definecolor{orl}{HTML}{EDBB99}
\definecolor{usecase}{HTML}{008FAC}
\definecolor{deepskyblue}{HTML}{00BFFF}
\definecolor{purple}{HTML}{C4A9FF}
\definecolor{bb}{HTML}{DAE3F3}
\definecolor{oo}{HTML}{FBE5D6}
\definecolor{gry}{HTML}{EDEDED}
\definecolor{yl}{HTML}{FFC000}
\definecolor{yll}{HTML}{FFFFE0}

\usepackage{tikz}
\newcommand*\circled[1]{\tikz[baseline=(char.base)]{
            \node[shape=circle,draw,inner sep=2pt] (char) {#1};}}

\definecolor{green}{rgb}{0.1,0.1,0.1}

\newcommand{\tzi}[1]{\textcolor{red}{FIX: #1}\xspace}

\usepackage{diagbox}
\usepackage{fontawesome5}

\usepackage{caption}
\usepackage{dcolumn, array, booktabs}
\newcolumntype{.}{D{.}{.}{-1}}

\makeatletter

\newcommand{\scriptveryshortarrow}[1][3pt]{{%
    \hbox{\rule[\scriptratio\dimexpr\fontdimen22\textfont2-.2pt\relax]
               {\scriptratio\dimexpr#1\relax}{\scriptratio\dimexpr.4pt\relax}}%
   \mkern-4mu\hbox{\let\f@size\sf@size\usefont{U}{lasy}{m}{n}\symbol{41}}}}

\makeatother

\newcommand{\acc}{\texttt{Accuracy}\xspace}
\newcommand{\sobol}{\texttt{Sobol}\xspace}
\newcommand{\shm}{\texttt{Shmembench}\xspace}
\newcommand{\cpy}{\texttt{Copy}\xspace}
\newcommand{\add}{\texttt{Add}\xspace}
\newcommand{\mul}{\texttt{Mult}\xspace}
\newcommand{\triad}{\texttt{Triad}\xspace}
\newcommand{\dott}{\texttt{Dot}\xspace}

\AtBeginDocument{%
  }

\newcommand{\sysname}{\textsc{Opal}\xspace}
\begin{document}

\title{\sysname: A Modular Framework for Optimizing Performance using Analytics and LLMs}


\author{Mohammad Zaeed}
\affiliation{%
  \institution{Texas State universitry}
  \city{San Marcos}
  \country{USA}}
\email{cup7@txstate.edu}

\author{Dr. Tanzima Z. Islam}
\affiliation{%
  \institution{Texas State universitry}
  \city{San Marcos}
  \country{USA}}
\email{tanzima@txstate.edu}

\author{Vladimir Inđić}
\affiliation{%
  \institution{University of Novi Sad}
  \city{Novi Sad}
  \country{Serbia}}
\email{vladaindjic@uns.ac.rs}






\begin{abstract}
Large Language Models (LLMs) show promise for automated code optimization but struggle without performance context. This work introduces \sysname, a modular framework that connects performance analytics insights with the vast body of published literature by guiding LLMs to generate informed, trustworthy optimizations. Unlike traditional performance tools that identify bottlenecks but stop short of actionable suggestions, \sysname bridges this long-standing gap by linking dynamic insights—from hardware counters and Roofline analysis to stall events—to optimization decisions. We evaluate \sysname across 1640 experiments on real-world GPU kernels and find that in over 98.5\% of cases, even a single insight source yields speedups, ranging on average from 19.34\% to 52.3\%. Our prompt template produced correct code in all but one case, where a vague diagnostic caused an unsafe suggestion. By automatically optimizing GPU kernels using performance analytics and LLMs, \sysname marks a leap toward democratizing expert-level performance engineering for all.
\end{abstract}




\maketitle

\section{Introduction}
\label{sec:intro}
As \gls{hpc} systems grow in complexity, writing high-performance GPU code demands deep architectural insight and expert-level interpretation of performance diagnostics. Profilers and models can pinpoint bottlenecks—but translating those diagnostics into code changes remains a manual, expertise-driven task. This final step creates a steep barrier: few developers have access to vendor engineers or system architects. 
Most are left staring at myriads of diagnostic messages and visualizations with no clear path to an actionable step. 

The rise of~\glspl{llm} such as GPT-4o~\cite{hesham2024fine} and LLaMoCo~\cite{ma2024llamocoinstructiontuninglarge} has created new opportunities for code optimization. Several efforts, such as DeepDev-PERF~\cite{garg2022deepdev} and RapGen~\cite{garg2023rapgen}, fine-tune LLMs for domain-specific tasks. However, most rely on static analysis~\cite{cummins2025llm,li2024enhancing} or prompt tuning~\cite{madaan2023learning,brown2020language,wei2022chain}, and ignore how the code performs on actual hardware. This is a big gap as optimization decisions are sensitive to runtime behavior---e.g., stalls, warp occupancy, memory throughput---and static inputs miss these signals. 
This gap between runtime diagnostics and targeted optimization motivates the central research question of this work: \textbf{Can LLMs, when guided by dynamic performance insights, generate valid and effective code transformations that directly address runtime bottlenecks?}
This gap is empirically validated in our experiments: when no performance insight is provided, the LLM fails to generate any optimizations (see Section~\ref{sec:defaultconfig}), confirming that performance context is essential for meaningful code transformation.

To bridge the gap, we present \sysname\footnote{\sysname stands for \textbf{O}ptimizing \textbf{P}erformance using \textbf{A}nalytics and \textbf{L}LMs}, a modular and extendable framework for automatically translating \textit{runtime diagnostics into optimizations}.
Unlike prior work that treats LLMs as black-box generators,~\sysname aims to leverage their reasoning capability. To achieve that, \sysname encodes multiple complementary performance diagnostics in a prompt described in natural language. 
We hypothesize that the LLM’s pretraining on HPC literature enables it to reason about code transformations that directly address runtime bottlenecks. To our knowledge, \sysname is the first framework to close the gap between diagnosis and transformation.

\sysname integrates three complementary sources of performance diagnostics: Roofline Analysis, Program Counter (PC) Stall Sampling, and 
Application–hardware interactions, as none of the sources alone is sufficient.
For instance, while counters quantify pressure on architectural components, they lack code-level precision; conversely, PC Sampling pinpoints code lines but 
cannot explain architectural boundedness.
Thus, these sources form a complementary triad: counters explain \textit{why} performance degrades across many configurations, Roofline analysis shows \textit{how} the kernel behaves on the architectural spectrum, and stall events pinpoint \textit{what and where} micro-level behaviors limit efficiency.
To leverage~\sysname, a user only needs to select the code and their choice of the diagnostic source. They can select none, one, two, or all three of the sources. 
Our extensive ablation study (Section~\ref{sec:defaultconfig}) demonstrates that \sysname can recommend code transformations with just a single source of information, reducing the average execution time by up to 64.93\%.

Building \sysname is not as simple as wrapping performance data into a prompt and querying an \gls{llm}. Dynamic performance profiles are often large—hundreds of megabytes—making it impractical to send raw data. \sysname addresses this by extracting only the most salient signals from each source, significantly reducing input size while preserving relevance.
Even with compact prompts, model edits are untrustworthy unless the reasoning behind them is clear. Without understanding why a change was made or how it links to diagnostics, users cannot validate or trust the transformation. To make reasoning transparent, \sysname uses belief tracing, which extracts token-level log-probabilities and reconstructs human-readable phrases by merging subword tokens based on a reference dictionary and a greedy sliding window.
Finally, \sysname presents users with the complete diagnostic-to-edit chain, including the original and modified code, the rationale behind each change, any deferred suggestions, and the specific performance bottleneck that motivated the transformation.
When a transformation is deemed unsafe, the~\gls{llm} does not apply it but still recommends it with justification based on the diagnostics.
These suggestions can serve as educational material for users or guide experts.
To ensure reproducibility, we use low-temperature decoding (0--0.15). 
Although this paper focuses on GPU kernels, the methodology 
applies to CPUs and other accelerators. \sysname can be easily extended to accommodate new sources and formats. Finally, while~\sysname currently uses a general-purpose~\gls{llm} in the backend, users can substitute local or specialized models of their choice without any modification to the code--making \sysname fully extensible and adaptable to evolving capabilities.

In summary, in this paper, we:
\begin{itemize}
    \item \textbf{Introduce} \sysname, a multi-source performance analytics-powered automated code optimization framework using~\gls{llms}.
    \item \textbf{Construct} a token-aware prompt that yields responses that link each optimization to its diagnostic source.

    \item \textbf{Design} a method to infer the reasoning of an \gls{llm} to identify what influenced each decision.
    \item \textbf{Validate} \sysname using extensive experiments across eight benchmarks on both NVIDIA and AMD GPUs, showing correctness and performance improvement, and explaining reasoning.
    \item \textbf{Implement} an end-to-end user interface for practitioners to explore, apply, and trust LLM-based optimizations, which will be made publicly available along with all data collected.
\end{itemize}

We also demonstrate that \sysname's recommended optimizations to two \gls{hpc} proxy applications---XSBench~\cite{tramm2014xsbench} and SW4lite~\cite{wu2021performance}. Moreover, to demonstrate that \sysname is not tied to a specific programming model, we select benchmarks with both CUDA and HIP implementations. Our study includes 8 real-world GPU applications,\footnote{Applications part of Babelstream~\cite{babel2019}, HeCBench\cite{jin2023a} and ECP proxy application suite~\cite{wu2021performance,tramm2014xsbench}.} evaluated across 68 input configurations and over 1640 experiments. These applications span a range of runtime bottlenecks, including memory-bound and compute-bound kernels. Despite these differences, all but one generated version passed benchmark-supplied correctness tests. While the exact transformations may vary between platforms, the diagnostic reasoning pipeline remains unchanged. Each optimization is generated in a few seconds and delivers up to 87.6\% reduction in execution time, with a 33.3\% average improvement across all trials. These results validate our core claim:~\sysname provides a generalizable, architecture-agnostic methodology for optimizing performance using analytics and LLMs. 

\textbf{Limitation}
To keep validation tractable, we synthesize optimizations from a single representative configuration per application. Although performance bottlenecks may vary across inputs, our experiments across 68 configurations show that \sysname's recommendations generalize well when performance characteristics remain consistent. This paper builds and validates the foundation for broader support; in future work, we plan to cluster input regimes based on shared bottleneck patterns and apply \sysname to a representative from each regime.

The rest of the paper is organized as follows: Section~\ref{sec:background} presents high-level information about each of the analytics sources; 
Section~\ref{sec:approach} describes our approach including salient information identification, prompt template, and belief tracing in details; 
Section~\ref{sec:framework} describes the implementation-specific details along with how users can use \sysname;
Section~\ref{sec:setup} describes the benchmarks and experimental setup in details; Section~\ref{sec:results} presents thorough evaluation. 
Finally, Section~\ref{sec:related-work} presents related literature and Section~\ref{sec:conclusions} concludes. 

\section{Background}
\label{sec:background}



\subsection{Roofline Analysis}
\label{sec:roofline}
To design effective optimization, it is required to know \textit{how} an application behaves: whether it is compute-bound or memory-bound, for instance. 
By comparing achieved performance against arithmetic intensity, Roofline models can reveal fundamental performance bottlenecks~\cite{williams2009roofline}--critical information not easily inferred from isolated counters or stalls. 
%
Several prior works have leveraged Roofline analysis to guide manual GPU optimizations. Yang et al.~\cite{yang2020hierarchical} used hierarchical Roofline modeling to identify bandwidth-bound kernels and applied loop tiling for better L2 cache locality. 
Bauer et al.~\cite{bauer2011cudadma} applied Roofline-based insights to recommend texture memory and warp-specialized data movement. Lee et al.~\cite{lee2020roofline} proposed a Roofline-based data migration strategy for hybrid memories and Williams et al.~\cite{williams2009roofline} used structure-of-arrays layouts to improve coalescing. On AMD GPUs, Leinhauser et al.~\cite{leinhauser2022metrics} used instruction-level Roofline analysis to guide LDS usage and memory coalescing. However, users need to be experts to infer which code transformation to apply given these insights. 

\subsection{PC Sampling-Based Analysis (PC)}
\label{sec:pc-sampling}

Program Counter (PC) Sampling provides fine-grained insights into \textit{where} in the source code performance stalls occur and \textit{what} microarchitectural reasons cause them. Unlike Roofline models, PC Sampling attributes specific stall types (e.g., \texttt{stall\_barrier}, \texttt{stall\_memory\_throttle}) to individual assembly instructions and their corresponding source lines. This precision has enabled tools such as GPU Performance Advisor~\cite{zhou2021gpa} to trace 40\% of stalls in a kernel to a single load and recommend targeted fixes. 
However, existing LLM-based tools such as~\cite{chen2021evaluatinglargelanguagemodels,zhou2024star} do not integrate such instruction-level data into their reasoning.~\sysname fills this gap by injecting these stall names, their meanings and specific code in the identified line into the structured prompt described in Section~\ref{sec:prompt} to guide the LLM to reason about micro-level bottlenecks. Still, PC Sampling data has limitations as it may miss rare stalls or input-dependent behaviors. This is why we augment these insights further using input-aware behavioral profiling via hardware counters (Section~\ref{sec:importance_extraction}).

\subsection{Application-Hardware Interaction Analysis (IA)}
\label{sec:importance_background}

While Roofline analysis highlights whether a kernel is compute- or memory-bound at a high level~\cite{williams2009roofline}, and PC Sampling pinpoints localized stalls at specific instruction lines~\cite{zhou2021gpa}, neither fully explains how performance is shaped by the interaction between the application and the architecture across varying configurations. This missing layer—application-specific, dynamic characterization—can only be captured by observing which hardware events consistently correlate with runtime degradation across multiple input sizes, thread block shapes, and control parameters.
\sysname addresses this gap by identifying application-hardware interactions that can explain the degradation. Hardware counters such as warp divergence, L2 cache misses, shared memory bank conflicts are critical for understanding why an application performs poorly. While prior work has shown their utility for isolated diagnosis~\cite{jensen2021filcio,hao2023drgpu,banchelli2023}, the challenge lies in automatically identifying which counters matter most, among hundreds exposed by modern architectures.


\section{Our Approach}
\label{sec:approach}
\begin{figure}
    \centering
    \includegraphics[width=\columnwidth]{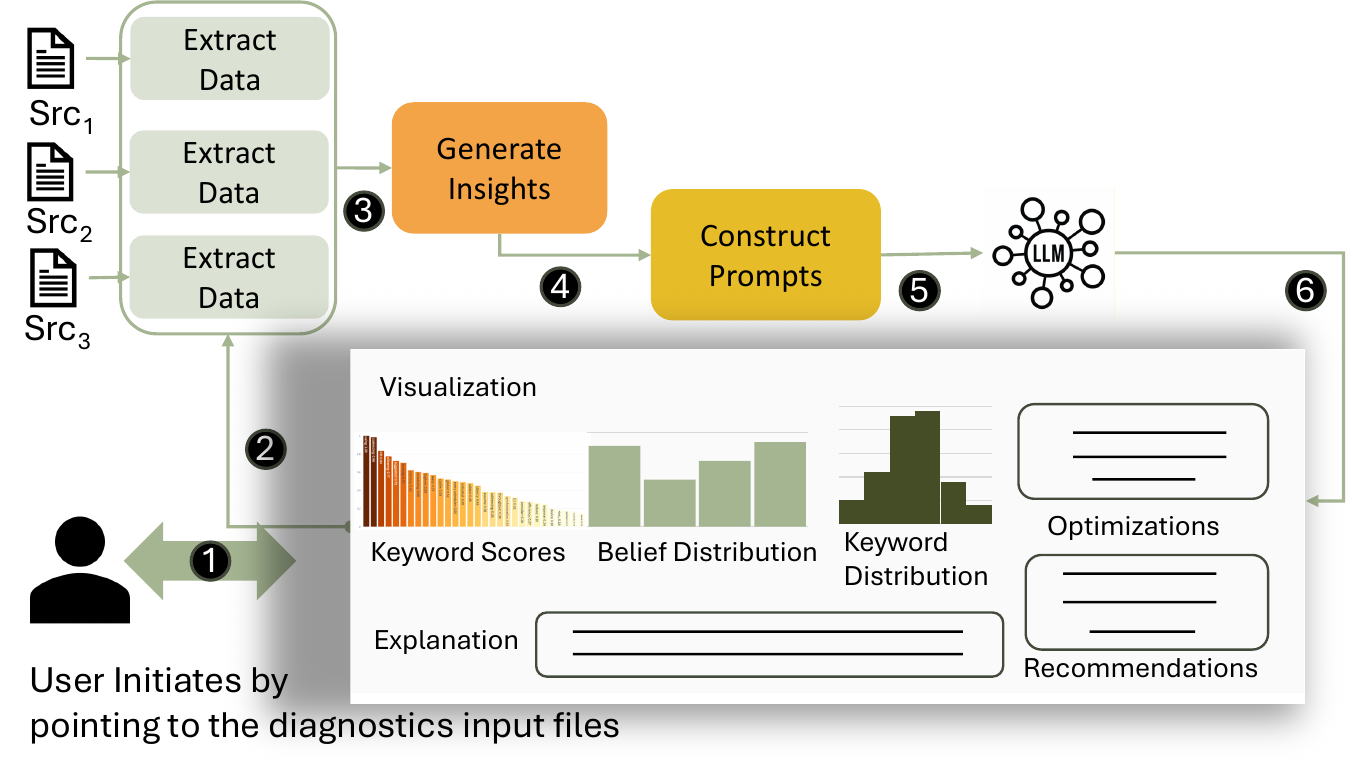}
    \caption{Overview of the~\sysname framework. Users select performance profiles, and \sysname constructs structured prompts compatible with any \gls{llm}, including custom models.}
    \label{fig:overview}
\end{figure}

%
Figure~\ref{fig:overview} shows the entire methodology of \sysname. The tasks can be decomposed into three parts: (1) generating important insights from different sources, (2) constructing prompts, (3) explaining~\gls{llm}'s reasoning using belief tracing. 


\subsection{Scope}
\sysname is designed to optimize individual GPU kernels. This scope aligns with prior work showing that kernel-level modeling is effective for isolating bottlenecks using Roofline analysis, hardware counters, and PC Sampling~\cite{williams2009roofline, yang2020hierarchical, leinhauser2022metrics, zigon2020utilizing, islam2025data}. A study of 31 GPU workloads found that 70\% of total execution time is spent in a single kernel~\cite{eeckhout2021cactus}, and commonly used profilers such as NVIDIA’s Nsight Compute~\cite{nvidia2023nsight} and AMD’s ROCm profiler~\cite{amd2023rocm} provide diagnostics at this level. Focusing on kernels ensures that optimizations remain interpretable and actionable.
\sysname currently supports performance profiles; extending support to handle time-series performance traces will require modifications to the \texttt{Extract Data} and \texttt{Generate Insight} modules in Figure~\ref{fig:overview}.
\subsection{Extract Data \& Generate Insights}
\label{sec:generate-insights}
Since the size of the raw data generated by the runtime diagnostic tools can be hundreds of megabytes in size, a naive approach of simply sending all data to an~\gls{llm} is impractical due to the limited token available for communication. This is why, \sysname implements source-specific salient information extraction methodology for effectively communicate critical information in token-efficient manner. \texttt{\sysname} 
prioritizes those performance problems with associated speedup suggestions or architectural bottlenecks, and integrates them into their respective \texttt{<$*$\_summary>} sections of the prompt template in Listing included in~\ref{lst:config}.

\subsubsection{Extract Insights from Roofline Data}
\label{sec:roofline-data-processing}
The \texttt{<roofline\_analysis\_summary>} section is populated differently for each architecture due to variations in vendor tooling. For NVIDIA GPUs, \texttt{\sysname} leverages the Roofline insights provided by Nsight Compute. These insights include both kernel-level metrics and accompanying diagnostic comments generated by the tool itself. For example, if Nsight Compute identifies that ``this kernel exhibits low compute throughput and memory bandwidth utilization relative to the peak performance of this device," it may recommend checking scheduler or warp state statistics. \texttt{\sysname} summarizes such rule-based diagnostics, prioritizes those with associated speedup suggestions or architectural bottlenecks, and integrates them into the \texttt{<roofline\_analysis\_summary>} section. This makes the prompt contextually rich and actionable without requiring the LLM to re-interpret raw numerical outputs.

In contrast, AMD Roofline reports are minimal and provide only numerical bandwidth and FLOPs data across memory levels (e.g., L1, L2, HBM, LDS) and operation types (e.g., FP32, FP64, MFMA). Lacking interpretive guidance, \sysname (1) identifies underutilized components by comparing observed and peak values, and (2) supplements this with insights from AMD optimization guides and modeling studies~\cite{amd_roofline_micro2022}. This enables the prompt to convey directional performance cues, even without automated analysis.

\subsubsection{Extract Insights from Stall Information}
\label{sec:pc-data-processing}
\label{sec:pc}
To pinpoint where in the application code performance bottlenecks manifest, \sysname integrates program counter (PC) Sampling data. While hardware counters explain what architectural inefficiencies exist and why they arise, PC Sampling identifies where these issues occur in the source code, enabling more actionable and localized optimizations.
\sysname analyzes assembly instruction-level samples collected via Nsight Systems and maps them to source-level line numbers using debug symbols. For each kernel, it identifies lines where individual stalls exceed a 10\% threshold and attributes those stall causes (e.g., memory dependency, execution dependency, pipeline busy).
Rather than directly including PC Sampling output, \sysname summarizes the root stall occurrences, line numbers, and adds that to the prompt in the designated $<$PC\_sampling\_summary$>$ section. For instance, raw PC Sampling output for \acc (41 lines of code) is 990 MB. After pre-processing, the data size reduces to 10 MB, and \sysname's summarization compresses it further to under 80 lines (6 KB). This 99.9\% reduction preserves actionable stall insights while being under \gls{llm}'s token limit.

\subsubsection{Extract Insights from Hardware Counters}
\label{sec:importance_extraction}
Modern architectures expose hundreds of hardware performance counters. Many are redundant, correlated, or irrelevant to optimization. 
Since in this study, our goal is to validate whether important signals exist across any subset, we collect the full set of available counters to avoid missing potentially useful ones. However, in practice, users do not need to collect every counter. If they already know a relevant subset based on domain knowledge or architectural experience,~\sysname can still operate on that list and systematically reduce it further to just the counters that can explain how performance changes across configurations. 

We frame this filtering task as a sparse feature selection problem. Each run of a kernel produces a vector of runtime measurements and a corresponding vector of hardware counter values. Together, they form a matrix $\mathbf{D} \in \mathbb{R}^{N \times C}$, where $N$ is the number of runs and $C$ is the number of counters, and a target vector $\mathbf{t} \in \mathbb{R}^N$ representing the runtime or any performance metric. We solve for a sparse weight vector $\mathbf{a}$ that identifies a small set of counters that best explain $\mathbf{t}$ using the formation in Equation~\ref{eqn:spcode}. To avoid overfitting and isolate only the most informative counters, we solve for a sparse weight vector $\mathbf{a}$ that minimizes the prediction error between observed performance $\mathbf{t}$ and the counter matrix $\mathbf{D}$. The sparsity level is controlled by $\kappa$, which limits the number of non-zero entries in $\mathbf{a}$ (Equation~\ref{eqn:spcode}). Higher sparsity causes a reduction in the number of insights provided to the \gls{llm}.
\begin{equation}
\min_{\mathbf{a}} \|\mathbf{t} - \mathbf{D}\mathbf{a}\|_2^2 \quad \text{s.t. } \|\mathbf{a}\|_0 \leq \kappa
\label{eqn:spcode}
\end{equation}
Because this problem is NP-hard, we use Ensemble Orthogonal Matching Pursuit (OMP), a randomized greedy algorithm. At each step, instead of picking just the most correlated counter, Ensemble OMP selects the top $\tau$ candidates, samples one at random based on their correlation strengths, and repeats. This ensemble approach explores multiple plausible solutions and avoids the local bias of standard greedy methods.

Finally, \sysname selects the top-$k$ (this work uses $k=5$) influential counters and summarizes their meanings in natural language. In our experiments, we observed no significant difference in performance outcomes between using the top-3 versus top-5 counters, so we chose $k=5$ to balance completeness with brevity.
These descriptions, generated once per architecture offline by mining vendor documentation, are stored in a reusable dictionary. The top-$k$ summaries are then embedded into the prompt to provide context, e.g.: ``\texttt{l1tex\_\_data\_bank\_conflicts} tracks shared memory bank conflicts, typically caused by misaligned or reduction operations, increasing latency."
For this step, we build on the publicly available Dashing repository~\cite{islam2019toward,dashing} and 
will make the dictionaries for NVIDIA and AMD GPUs publicly available.

\subsection{Prompt Construction}
\label{sec:prompt}

To ensure consistent and parseable responses,~\sysname uses a prompt template with programmatically populated placeholders. As shown in Listing included in~\ref{lst:config}, each prompt begins with the unoptimized kernel code annotated with line numbers. This section is followed by structured summaries from each diagnostic source. A final instruction block then directs the~\gls{llm} to (1) generate optimized code, (2) explain each change by linking it to a specific diagnostic insight, and (3) list additional optimizations that were considered but not applied, along with the corresponding evidence. These deferred suggestions serve as actionable leads for human experts. \sysname parses the response and presents the optimizations and suggestions along with their diagnostic rationale to user through an interactive user interface.








\noindent\textbf{Listing\,1:} The prompt template used by \texttt{\sysname}
\label{lst:config}
{\footnotesize\ttfamily
\begin{verbatim}
code:
  snippet: "<source_code_with_line_numbers>"

roofline_summary: <roofline analysis summary>

stall_analysis_summary: <stall analysis summary>

key_hardware_events: <counter analysis summary>

task_instruction: You are an HPC performance optimization expert. Optimize 
    the provided <programming language> code specifically for <architecture> 
    on input configuration <input config>. Clearly reference the diagnostic
    data by number or quoted text in each inline comment.

Required Output Format:

1. Provide the complete optimized code wrapped within ```cpp code blocks```.
2. After the code, list applied changes:
    optimizations = [ {'lines': [line_numbers], 'reason': 'justification from 
    diagnostic'}, ... ]
3. List suggestions not applied: suggested_but_not_applied = 
    [ {'lines': [line_numbers], 'reason': 'diagnostic, but deferred due to 
    uncertainty'}, ... ] 
4. Do not include any additional explanation beyond the code and the two lists.
\end{verbatim}
}

\subsection{Explain LLM's Reasoning using Belief Tracing}
\label{sec:logprobe}
To explain why \sysname suggests a particular code transformation, users need visibility into the model’s internal reasoning. 
\gls{llms} produce output one token at a time by selecting the most likely next token based on prior context.
Internally, the model assigns a probability $P(x)$ to every possible next token $x$ in its vocabulary, and selects the one with the highest likelihood.
For instance, when generating an explanation about optimizing memory access in the \acc kernel, the model may generate a sequence of tokens that together form the phrase \texttt{memory coalescing}. This occurs because it has learned from the literature that ``memory" and ``coalescing" together describe a meaningful concept in GPU optimization.
However, models do not explicitly tell us which high-level phrases they considered important. Instead, they emit subword tokens such as ``co", ``alesc", and ``ing", without revealing how those pieces combine. 

\textbf{Calculate belief scores for \texttt{logprobes} tokens}
To provide explainability to \sysname's reasoning, we introduce a mechanism called belief tracing. Specifically, \sysname uses the \texttt{logprobes} API to extract the log-probability of each generated token. However, raw log-probabilities are difficult to interpret: they are negative, unbounded, and inversely correlated with confidence. 
Lower log-probability indicates lower confidence. To convert this signal into a more intuitive scale, we define a belief score \( B(x) \) for each token in Equation~\ref{eq:belief}:
\begin{equation}
\label{eq:belief}
B(x) = \exp(-\alpha \cdot \log P(x))
\end{equation}
where \( \alpha \) is a positive scaling constant (we use \( \alpha = 2 \)). This transformation maps the log-probabilities to the $[0, 1]$ range, where values closer to 1 indicate higher model confidence. It preserves the relative differences across tokens while inverting the log scale, so that less likely (more surprising) tokens receive lower belief scores. These belief scores form the basis for identifying which keywords the LLM focused on during code generation.

\textbf{Construct keywords from tokens}
Since the generated tokens are subwords, next, we construct keywords from these tokens by using a reference dictionary~$\mathcal{R}$ created from the input prompt, extracting all $n$-grams ($n \leq 4$).
We then scan the logprobe token sequence using a greedy sliding window, merging the longest contiguous span that appears in~$\mathcal{R}$. If a match is found, we assign the belief score of the phrase~$w = \{t_1, \ldots, t_n\}$ in Equation~\ref{eq:phrase}:
\begin{equation}
\label{eq:phrase}
B(w) = \prod_{i=1}^n B(t_i)
\end{equation}
This multiplicative aggregation captures compound reasoning--phrases are assigned high belief only if all component tokens are influential. If no match is found, we retain the original single-token belief. This reconstruction step thus calculates importance of domain-specific keywords in $\mathcal{R}$ for explanation. 

\textbf{Filtering}
Finally, we filter all punctuation, numeric fragments, or partial words and those: (1) with non-alphabetic characters only, (2) with length $\leq 2$ unless alphabetic, and (3) not in $\mathcal{R}$. Then, we apply logarithmic scaling to compress large values and amplify small differences 
by adding a small constant ($\epsilon = 10^{-7}$) to avoid undefined behavior for zero-valued beliefs. 
and re-normalized to \([0, 1]\) using min-max scaling. 
Section~\ref{sec:reason} shows an example of how these keywords and their beliefs explain reasoning.



\section{The \sysname Framework}
\label{sec:framework}

Figure~\ref{fig:overview} illustrates the end-to-end workflow of \sysname. Users start by selecting the source code and any number of diagnostic inputs via a lightweight Streamlit-based dashboard. Currently supported inputs include PC Sampling logs, hardware counters, and Roofline analysis, all in JSON format (\circled{1}). Data collection is performed offline using standard profiling tools commonly used in HPC workflows~\cite{williams2009roofline,HPCToolkit,islam2025data}.
After file selection, \sysname invokes a separate \texttt{Extract Data} module for each input source (\circled{2} in the figure). Each module parses source-specific outputs into structured format, as described in: (1) Section~\ref{sec:pc-data-processing} for PC Sampling logs, (2) Section~\ref{sec:roofline-data-processing} for Roofline analysis, and (3) Section~\ref{sec:importance_extraction} for hardware counter attribution.
These structures are passed to the \texttt{Generate Insights} module (Step \circled{3}), which summarizes and compresses each input into 
prompts. This insight reduction pipeline, detailed in Section~\ref{sec:generate-insights}, ensures the prompt remains actionable while avoiding verbosity. 

Next, the Construct Prompts module (Step \circled{4}) combines the unoptimized kernel and diagnostic summaries into a structured prompt following a template (Listing included in~\ref{lst:config}). This step, described in Section~\ref{sec:prompt}, 
links each performance issue to corresponding parts of the code.
The prompt is then sent to an \gls{llm} with log probability tracking enabled (temperature=0.15), as shown in Step \circled{5}. 
The LLM returns three outputs: (1) an optimized kernel, (2) explanations justifying each transformation, and (3) deferred suggestions that were considered but not applied due to uncertainty.

\subsection{User Interaction and Visualization}
\sysname displays the results through an interactive web dashboard (Step \circled{6}). Users can examine: (1) the optimized code, with each modification linked to its triggering diagnostic, (2) explanations for each change, including rationale for rejected transformations, and (3) visualizations that expose how the \sysname reached its decisions.
The visualization panel includes: (1) a flame-style Keyword Score bar chart showing belief scores for the most influential prompt tokens,
A Belief Distribution histogram indicating \sysname's overall confidence (calculated in Section~\ref{sec:logprobe}),
and a Keyword Distribution chart summarizing recurring reasoning patterns.
These are derived from belief tracing (Section~\ref{sec:logprobe}), which reconstructs meaningful phrases from token-level logprobs and highlights the reasoning path the model followed. Every decision is traceable—down to the diagnostic message and the specific lines of code affected—allowing users to validate, override, or extend the transformation.

\sysname is modular by design. To support new tools such as HPCToolkit~\cite{HPCToolkit}, users only need to extend the \texttt{Extract Data} module. All other components remain unchanged.
In future, \sysname can be integrated into a multi-agent AI framework, where agents autonomously run measurement tools, process the logs, and constructs the prompts in a collaborative manner. This will unify data collection and analysis.

\section{Evaluation Setup}
\label{sec:setup}
\begin{table}[t]
\footnotesize
\caption{Benchmark applications, their size, and tunable configurations.}
\label{tab:app-config-summary}
\begin{singlespacing}
\centering
\resizebox{\columnwidth}{!}{
\begin{tabular}{|p{1.5cm}|c|p{5.2cm}|}
\hline
\textbf{App.} & \textbf{LOC} & \textbf{Configuration Description} \\
\hline

\texttt{Copy} & 2 & \multirow{5}{5.2cm}{\parbox{5.2cm}{ARRAY\_SIZE$\in$[2000896,4000768,6000640], \texttt{num\_times}$\in$[100,300], step=100; \newline Default=(2000896,100).\\ARRAY\_SIZE: dataset size; \newline \texttt{num\_times}: repetition count.}} \\
\cline{1-2}

\texttt{Mul} & 3 &  \\
\cline{1-2}

\texttt{Add} & 2 &  \\
\cline{1-2}
\texttt{Triad} & 3 &  \\
\cline{1-2}

\texttt{Dot} & 19 & \\
\hline

Shmemench~\cite{meswani2012tools} & 35 & \parbox{5.2cm}{rep$\in$[300,1500], step=300; Default=600.\\\texttt{rep}: kernel repetition count.} \\
\hline

\acc~\cite{jin2023a} & 41 & \parbox{5.2cm}{(\texttt{nrows}, \texttt{ndims}, \texttt{top\_k}, \texttt{rep})=8192 $\times$ [5000,10000] $\times$[10,20] $\times$ [100,300]; Default=(8192,5000,10,100).\\\texttt{nrows}: prediction vectors; \texttt{ndims}: dimensions/vector; \texttt{top\_k}: threshold; \texttt{rep}: repetition count.} \\
\hline

\sobol~\cite{jin2023a} & 195 & \parbox{5.2cm}{(\texttt{n\_vectors}, \texttt{n\_dimensions}, \texttt{repeat})=[10K,1M] $\times$ [1K,10K] $\times$ 100; Default=(10K,1000,100).\\\texttt{n\_vectors}: vectors generated; \texttt{n\_dimensions}: dimensions/vector; \texttt{repeat}: repetition count.} \\
\hline

\texttt{addsgd4\_SM} (SW4Lite~\cite{wu2021performance}) & 149 & \parbox{5.2cm}{(\texttt{input\_datasets})= [pointsource, uni, uni\_rev, skinny, gaussianHill]; Default=(pointsource).} \\
\hline

XSBench~\cite{tramm2014xsbench} & 81 & \parbox{5.2cm}{(\texttt{s}, \texttt{G})=[small, large] $\times$ [unionized, nuclide, hash]; Default=(large, unionized).\\\texttt{s}: problem size; \texttt{G}: grid search type.} \\
\hline
\end{tabular}
}
\end{singlespacing}
\end{table}

\subsection{Hardware and Environment} We evaluate \sysname on two GPU platforms: (1) NVIDIA H100 GPUs and (2) AMD MI 210 on
an unnamed cluster (hidden for double blind review). Each system provides modern CPUs, high-bandwidth interconnects, and GPU-optimized software stacks. Experiments use CUDA-12.3, Nsight Compute-2025.1.1.0, ROCm-rocm/6.3.2, Omniperf\-2.0.1 and GPT-4o via the OpenAI API. Every configuration is run 10 times, and results report the mean with error bars.

\subsection{Benchmarks and Configurations} We use four GPU applications comprising eight kernels that stress distinct bottlenecks: memory-bound, compute-bound, and stall-sensitive behavior. Table~\ref{tab:app-config-summary} summarizes each application's LOC (lines of code), parameter space and default configurations. These benchmarks include simple kernels (\cpy, \mul, \add, \triad, \dott), \shm for shared-memory operations, the compute-intensive \acc kernel, and the memory-sensitive \sobol kernel. These diverse workloads ensure a thorough evaluation of each diagnostic source's impact on LLM-driven optimization. The chosen benchmarks specifically have both CUDA and HIP implementations, enabling a cross-platform validation of our methodology.

\subsection{Data Collection Tools}
\textbf{Roofline Analysis}
For NVIDIA GPUs, we use Nsight Compute (\texttt{ncu}) to collect Roofline metrics, including arithmetic intensity, throughput, and diagnostic messages:
\textcolor{blue}{\texttt{ncu ----csv ----import \newline <output\_file> ----page details > <roofline\_file>.csv}}. AMD GPUs provide numeric Roofline data (bandwidth, FLOPs) via rocprof, without interpretive diagnostics. We manually identify bottlenecks by comparing these metrics to theoretical peaks, supplemented by AMD optimization guides~\cite{amd_roofline_micro2022}.

\textbf{PC Sampling}
PC Sampling requires compiling the application with debug information enabled using either the \texttt{-G} or \texttt{-lineinfo} flag. PC Sampling support is currently unavailable for AMD GPUs but can be integrated once vendor support becomes available.

\textbf{Hardware Performance Counters}
We use NVIDIA’s Nsight Compute tool to collect all available counters using the full preset (\texttt{--set full}) flag using the following command:
\textcolor{blue}{\texttt{ncu ----target-processes all --set full -o <output\_file> ./<app> <args>}}
This allows users to capture the entire list of counters without selecting specific ones manually. Because this command is backend-agnostic with respect to CUDA versions, it supports forward compatibility. 
For AMD GPUs, similar counter data is collected using Omniperf~\cite{omniperf2024}. Advanced users may specify subsets to streamline data collection. \sysname then applies the eOMP algorithm to identify important counters (Section~\ref{sec:importance_extraction}).

\subsection{Baselines} We compare optimizations generated by \sysname across seven scenarios against the performance of the unmodified code. The scenarios include: (1) PC Sampling (PC), (2) Importance Analysis (IA), (3) Roofline, (4) PC Sampling with Importance Analysis (PC+IA), (5) PC Sampling with Roofline (PC+Roofline), (6) Importance Analysis with Roofline (IA+Roofline), and (7) PC Sampling with Importance Analysis and Roofline (PC+IA+Roofline).
Additionally, to demonstrate the benefit and applicability of \sysname to \gls{hpc}, we select two proxy applications---XSBench~\cite{tramm2014xsbench} and a kernel named \texttt{addsgd4\_SM} from the SW4Lite benchmark suite~\cite{wu2021performance}.
We do not compare with hand tuned baselines as they depend strongly on domain expertise. We also evaluated an additional baseline, simply uploading the code to ChatGPT-4o. However, under our minimal $temperature$ setting, the \sysname did not receive any optimizations, and thus, we exclude this baseline from our comparison.

\subsection{Metrics Collected}
We collect the following metrics per application and for a selected set of configurations to keep the number of experiments tractable. 
We measure performance improvements primarily using wall-clock execution time. To validate correctness, every optimized kernel is tested against benchmark-supplied validation tests. \textbf{Notably, 100\% of the LLM-generated code transformations passed these checks using our current prompts}. Additionally, we present detailed validations of Roofline insights, PC Sampling data, and key hardware counter metrics for the optimized versions of a selected set of kernel. We also analyze keyword belief scores to explain the \gls{llm}'s reasoning processes.

\section{Results}
\label{sec:results}

The primary goal of our evaluation is to test the hypothesis that integrating performance insights into structured prompts enables~\gls{llms} to produce code transformations that directly target observed bottlenecks and improve runtime. To assess this hypothesis, we:
\begin{itemize}
    \item \textbf{Quantify} the contribution of each insight source via ablation.
    \item \textbf{Evaluate} the effectiveness of code transformations across many configurations.
    \item  \textbf{Validate} that the bottlenecks actually get fixed. 
    \item \textbf{Generalize}~\sysname's approach to AMD GPUs.
    \item \textbf{Explain}~\sysname's reasoning.
    \item \textbf{Demonstrate} on \gls{hpc} proxy applications.
\end{itemize}

\subsection{Quantify the contribution of each insight source via ablation}
\begin{table}[t]
\footnotesize
\caption{Default configuration and corresponding optimization summary for each benchmark on NVIDIA H100 GPUs.}
\label{tab:default-opt-summary}
\begin{singlespacing}
\centering
\resizebox{\columnwidth}{!}{
\begin{tabular}{|p{1.2cm}|p{2cm}|p{5.2cm}|}
\hline
\textbf{App.} & \textbf{Default Config.} & \textbf{Optimization Summary} \\
\hline

\texttt{Copy} & (2000896, 100) & Replaced \texttt{cudaMemcpy} with \texttt{cudaDeviceSynchronize} for managed memory to reduce overhead. Added boundary checking clause. \\
\hline

\texttt{Mul} & (2000896, 100) & Same as Copy \\
\hline

\texttt{Add} & (2000896, 100) & Same as Copy \\
\hline

\texttt{Triad} & (2000896, 100) & Same as Copy \\
\hline

\texttt{Dot} & (2000896, 100) & Same as Copy \\
\hline

Shmemench & 600 & Simplified vector assignment (line 28) to reduce instruction count; removed \texttt{\_\_threadfence\_block()} comment (line 48) to reduce MIO pipeline stalls. \\
\hline

\acc & (8192,5000,10,100) & Added \texttt{\_\_restrict\_\_} to pointers to reduce uncoalesced loads and enable alias-free memory access optimization. \\
\hline

\sobol & (10K,1000,100) & Replaced float conversion with \texttt{\_\_uint2float\_rn} at lines 29 and 50 to reduce latency and improve throughput; increased block size to 128 at line 66 to improve occupancy and GPU utilization. \\
\hline

\texttt{addsgd4\_SM} & pointsource & Unrolled loop and removed array offset function calls (lines 12702–12722) to reduce register pressure and improve locality. \\
\hline

XSBench & (large, unionized) & 
Increased block size to 512 to improve occupancy (line 5); added shared memory to reduce global memory access (line 6); reorganized memory access for coalescing (line 7); reduced register usage to boost occupancy (line 8); used warp-level primitives to minimize divergence (line 9); replaced double with single precision for faster computation (line 10); unrolled loops to reduce overhead and increase throughput (line 11).\\
\hline

\end{tabular}
}
\end{singlespacing}
\end{table}

\label{sec:defaultconfig}
\begin{figure}[t]
    \centering
    \includegraphics[width=\columnwidth]{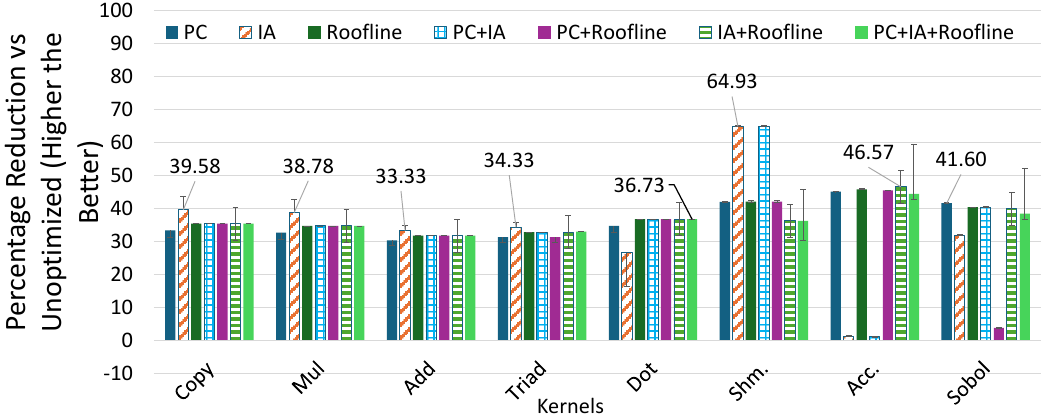}
    \caption{%
Ablation study of the impact of diagnostic sources on optimization performance. Error bars represent performance variability across repeated runs. IA alone achieves significant single-source gains, notably 64.93\% on \shm. PC+IA consistently delivers stable optimizations across kernels. \sobol shows substantial variability, indicating that the type of information can impact the effectiveness of code transformation. 
    }
    \label{fig:allapps-defaultconfig}
\end{figure}
The goal of this experiment is to quantify how much each performance insight contributes individually and in combination to guiding~\gls{llms} toward producing faster code.
Figure~\ref{fig:allapps-defaultconfig} shows results for eight GPU kernels on an NVIDIA H100, comparing optimizations generated using single sources, pairwise combinations, and all three sources combined. The X-axis lists applications, and the Y-axis shows the average percentage execution-time reduction relative to the unoptimized baseline, measured over 10 runs. Higher bars represent more effective \gls{llm}-guided optimizations. Error bars indicate variability: narrow bars imply consistent improvements, while wide bars highlight cases with significant fluctuations.

We restrict this study to one default configuration per application (defaults marked in Table~\ref{tab:app-config-summary}) to 
keep the number of experiments tractable.
Including all analysis sources across multiple configurations of multiple applications would exponentially increase the number of experiments. Section~\ref{sec:allconfig} investigates if recommended optimizations for the default configuration remains effectively for other configurations. Table~\ref{tab:default-opt-summary} summarizes all the optimizations automatically applied to the benchmark applications.

\textbf{Key observations} From Figure~\ref{fig:allapps-defaultconfig}, we observe that:
(1) Each source, when used independently, provides measurable average performance gains across all but one kernels (between $1.2 - 64.93\%$).
(2) Double source improvements range from $1.0 - 64.93\%$. 
(3) The full combination of all three sources (\texttt{PC + IA + Roofline}) achieves the most consistent gains across all applications. The average range varies from $31.8 - 44.49\%$. 
(4) Some of the optimizations cause run to run variability. Further investigation shows that 

\textbf{Not all optimizations are safe, guidance helps fix it} 
All but one generated code variants (out of 1640 experiments) passed benchmark-supplied validation checks for output correctness.
In that one case, \sysname incorrectly suggested \texttt{\_\_int\_as\_float} to reduce memory stalls, leading to test failure. The root cause was a vague counter description that failed to warn against unsafe casting. After refining the explanation to highlight bandwidth saturation risks and explicitly discourage this intrinsic, \sysname consistently chose the correct alternative, \texttt{\_\_uint2float\_rn}, eliminating correctness issues and runtime variability.
\subsubsection*{Effectiveness of Insight Types by Bottleneck Class}

Not all performance insight sources contribute equally; their usefulness depends on the nature of the application bottleneck. As shown in Figure~\ref{fig:allapps-defaultconfig}, the \texttt{accuracy} kernel benefits significantly from \textbf{Roofline analysis and PC Sampling}, each delivering over 45\% average execution time reduction. In contrast, IA alone yields negligible gain (1.2\%).

This contrast stems from the type of bottlenecks each source reveals. Roofline analysis identified key architectural inefficiencies in the \texttt{accuracy} kernel—underutilized compute pipelines and poor DRAM fetch granularity—while PC Sampling exposed persistent stalls such as \texttt{stall\_wait} and \texttt{stall\_membar}. Both of these signal memory-level throttling. The hardware counter IA, on the other hand, surfaced only three moderately weighted events, which primarily described shared memory usage and L1TEX-level store pressure—important but not dominant.

This pattern suggests that macro-level models (like Roofline) and line-level stall diagnostics (from PC Sampling) are more effective when memory coalescing, warp scheduling, or pipeline utilization is the core issue. In such cases, these sources guide \sysname to concrete, literature-supported transformations such as pointer alias removal or loop restructuring. We validate one such transformation in detail below.

\textbf{Why IA underperformed for \acc} Figure~\ref{fig:allapps-defaultconfig} shows that the benefit of using hardware counter analysis alone was only $1.2\%$ for \acc. This is because there were only three moderately important hardware events for \acc, primarily pointing to L1TEX store pressure and shared memory underuse. Addressing these would require invasive changes to data layout or memory hierarchy--transformations that are risky and often workload-specific. Hence, \sysname refrained from applying these transformations and instead provided guidance to a user. 



\subsection{Evaluate the effectiveness of optimizations across many configurations}
\label{sec:allconfig}
\begin{figure}[t]
    \centering
    \includegraphics[width=\columnwidth]{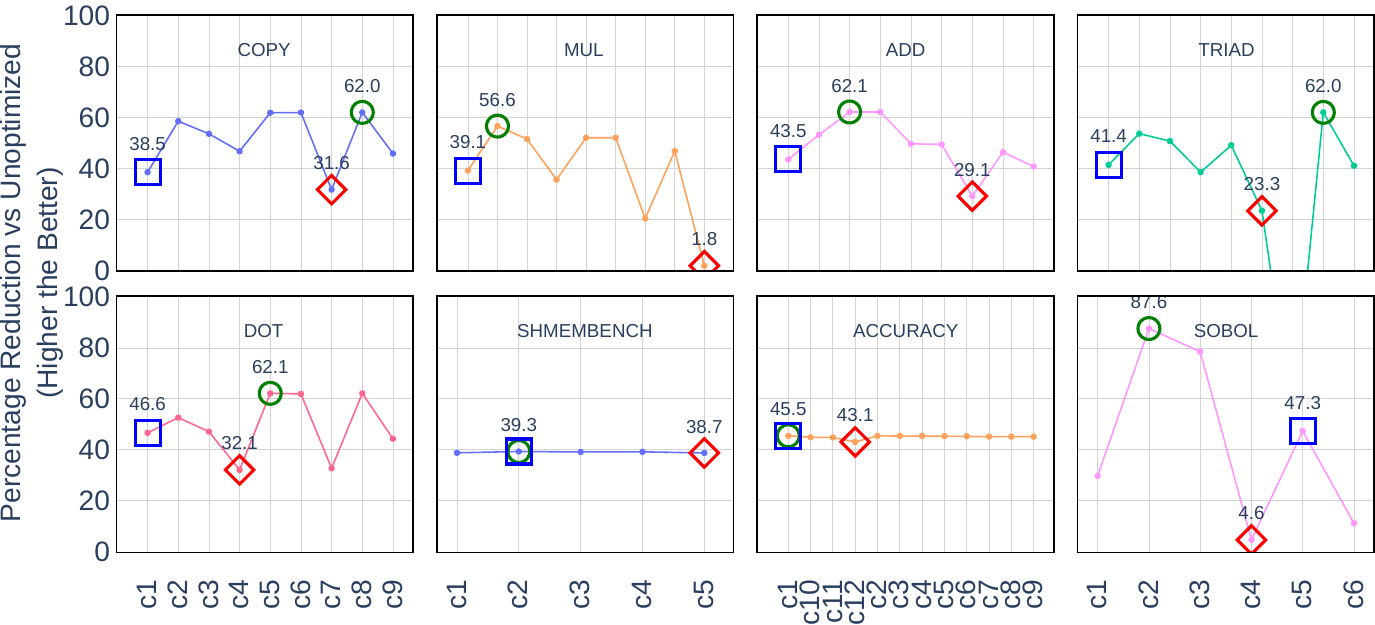}
    \caption{%
    Performance improvements across input configurations. 
    Blue rectangles represent improvements for default configuration (Table~\ref{tab:app-config-summary}), green circles indicate maximum improvement, and red diamonds indicate minimum improvement. This figure shows that optimizations generated for one configuration also yield performance improvements across other configurations, however, the margin varies across applications. \acc achieves the most stable gains (43.1\%–45.5\%), while that for \sobol varies widely (4.6\%–87.6\%).
    }
    \label{fig:allapps-allconfig}
\end{figure}
The objective of this experiment is to assess whether \sysname’s code transformation decided based on a single configuration generalize across diverse input sizes and configurations. As listed in Table~\ref{tab:default-opt-summary}, we evaluate eight kernels across 68 total configurations. Each run compares the unoptimized version to the \sysname-optimized version generated using analytics from the default configuration. Figure~\ref{fig:allapps-allconfig} plots the percentage reduction in execution time: the X-axis shows each configuration; the Y-axis shows the percentage reduction in runtime relative to the unoptimized baseline.

\textbf{Key observations} From Figure~\ref{fig:allapps-allconfig}, we can observe that: (1) reduction in execution time persist across all but 1 out of 68 configurations across 8 applications, demonstrating an overall generalizability of the observation that multi-source analytics insights-guided code transformations suggested by \sysname can be a great starting point and a lot better than starting from scratch. (2) The only configuration for \triad that achieved no benefit was when the kernel ran with an array size of 6000640 and 100 repetitions of kernels. 
(3) Kernels like \shm (up to 69.10\%) and \sobol (up to 87.60\%) see the largest gains, likely due to recurring bottlenecks across input ranges that let~\sysname generalize effectively. However, the improvements for \sobol drop for $C4$, respectively, suggesting input-sensitive behavior where bottlenecks shift with configuration. (4) In contrast, \acc consistently improves (43.1\%–45.5\%) across all inputs. The applied transformation (\texttt{\_\_restrict\_\_}) targets aliasing-related stalls that persist regardless of scale, indicating stable bottlenecks and validating that optimizations from one configuration can transfer when performance characteristics remain consistent.

\subsection{Validate whether post-optimization performance aligns with \sysname's stated rationale}

The objective of this experiment is to assess whether \sysname's code transformations mitigate the performance bottlenecks identified in the analysis phase. To keep the validation tractable, we focus on the two largest kernels, \acc and \sobol, using their default configurations. We re-run all analysis on both unoptimized and optimized versions to examine changes in key performance characteristics. These kernels expose different types of bottlenecks, making them ideal for evaluating whether \sysname's choices align with known optimization strategies. Validation is considered successful if the problematic behaviors reduce and if the explanations are consistent with published literature.

\subsubsection{Validating Bottleneck Elimination in \acc}
\begin{figure}[t]
    \centering    \includegraphics[height=0.3\textheight,
    keepaspectratio]{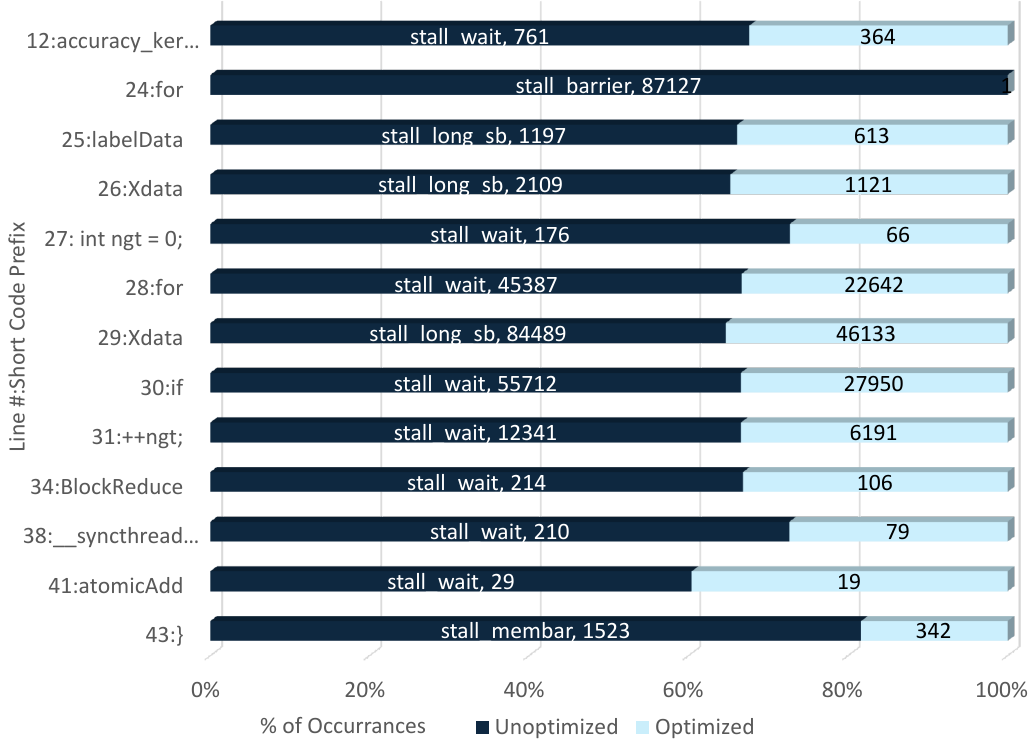}
    \caption{%
    Comparison of stall occurrences between unoptimized (left) and optimized (right) \acc kernels. The Y-axis lists line numbers and shortened code snippets. Each bar is labeled by the dominant stall type. 
    At line 6, \texttt{stall\_wait} occurrences decrease from $45,387 \to 22,642$; similar significant reductions occur at lines $29 \to 31$, validating the effectiveness of targeted optimizations.
    }
    \label{fig:accuracy_pcsampling_comparison}
\end{figure}

Performance profiling revealed poor warp readiness (0.16 eligible warps per cycle), inefficient DRAM fetch (2.7 sectors per 128-byte cache line), and persistent memory stalls, including \texttt{stall\_barrier} and \texttt{stall\_membar}. To address these, \sysname added the \texttt{\_\_restrict\_\_} keyword to \texttt{Xdata} and \texttt{labelData}. This keyword disambiguates pointer aliasing, which makes the compiler safely reorder memory accesses and apply coalescing optimizations. Published literature confirms that this keyword 
is used to enable contiguous memory access, which reduces transactional overhead~\cite{cuda_guide2024,williams2009roofline,crago2018exposing}. The post-optimization metrics in Figure~\ref{fig:accuracy_pcsampling_comparison} show that (1) scheduler efficiency improves from one instruction every 6.4 cycles to 3.5, (2) eligible warp rate increases from 0.16 to 0.40, and (3) DRAM fetch efficiency improves as indicated by a lower L2 miss-to-sector ratio. These improvements confirm that the \gls{llm} correctly identified a transformation 
that aligns with expert practices--validating that \sysname's reasoning is both explainable and effective.

\subsubsection{Validating Bottleneck Elimination in \sobol}

Similarly, we also validate the same for \sobol. For \sobol, the PC Sampling data identified three issues: low occupancy, MIO stalls linked to shared memory, and expensive float conversion operations marked by high \texttt{stall\_math\_ratio}. In response,~\sysname replaced \texttt{(float)X} with the intrinsic \texttt{\_\_uint2float\_rn(X)} in two critical lines. This transformation is 
supported by CUDA literature advocating the use of intrinsics to bypass normalization overhead. Post-optimization metrics 
(not shown) confirm the intended effect: occupancy increases from 25.6\% to 48.4\%, scheduler throughput improves from 1.9 to 1.6 cycles/issue, and stall types shift from MIO to fixed-latency dependencies--typical of highly optimized code. These changes validate that the LLM's recommendation was not only guided by real profiling data, but also converged on an established best-practice transformation shown to reduce instruction latency in throughput-bound kernels~\cite{cuda_guide2024, hijma2023optimization}. 

\subsection{Generalize~\sysname's Approach to AMD GPUs}

\begin{figure}[t]
    \centering
    \includegraphics[width=0.48\textwidth]{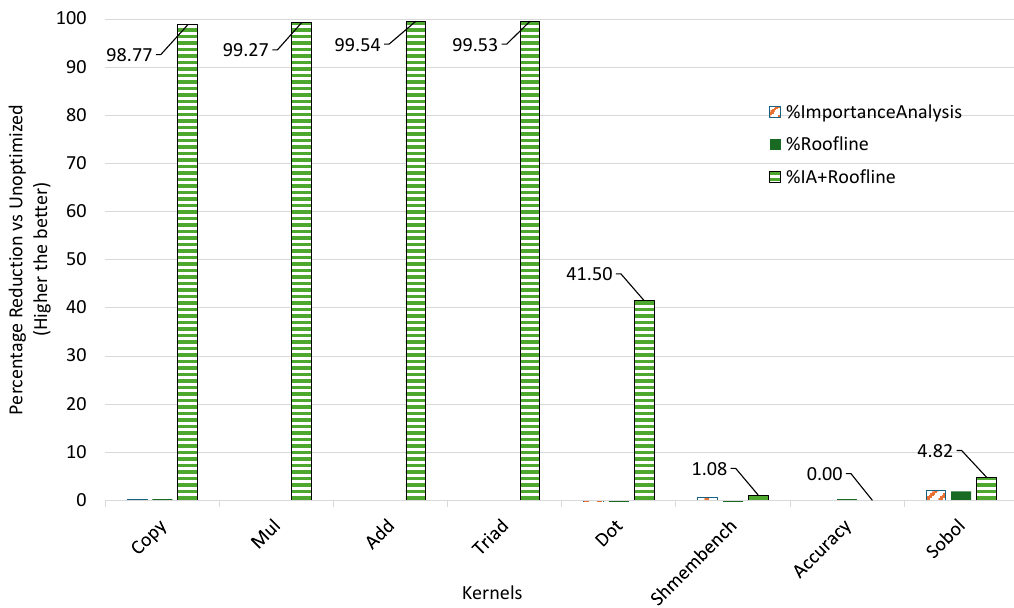}
    \caption{%
    \textbf{Ablation study on source contributions to optimization performance for HIP applications.} 
    We observe massive improvements for Babelstream kernels (41.5 to 99.27\%) using IA and Roofline analysis together as performance insights. For all other kernels, there is no improvement (some improvements are shown, but that is purely because of GPU execution uncertainty). 
    }
    \label{fig:amd-allapps-defaultconfig}
\end{figure}
The objective of this experiment is to assess whether \sysname generalizes to HIP applications and AMD GPUs. We replicate the ablation study from our CUDA evaluation using HIP versions of the same kernels. Performance data is collected on AMD MI210 GPUs using Omniperf. Due to tool limitations, PC Sampling is currently unsupported, and Roofline analysis lacks diagnostic commentary. Instead, we extract raw Roofline metrics—including \textbf{HBM Bandwidth}, \textbf{L2 Bandwidth}, \textbf{L1 Bandwidth}, \textbf{LDS Bandwidth}, and peak compute throughput across \textbf{FP32 VALU}, \textbf{FP32 MFMA}, \textbf{FP16 MFMA}, and \textbf{INT8 MFMA}. These values are passed into the prompt using the same structured format as in the CUDA pipeline.

\textbf{Key observations} From Figure~\ref{fig:amd-allapps-defaultconfig}, we observe:
(1) Single-source prompts (e.g., Roofline-only or counter-only) fail to yield performance gains across all HIP kernels. This is because Omniperf outputs only raw numeric values without descriptive bottleneck summaries. In the absence of contextual information—such as whether a kernel is memory- or compute-bound—the LLM struggles to reason about actionable optimizations. Unlike Nsight Compute, which provides direct language-level insights (e.g., ``kernel underutilizing SM"), Omniperf requires the LLM to infer everything from sparse numbers, which limits its effectiveness.

(2) In contrast, combining both Roofline and counter-based insights leads to substantial improvements in \cpy (41.50\%), \mul (78.43\%), \add (93.00\%), \triad (99.54\%), and \dott (86.31\%). These improvements are largely due to a consistent LLM-generated fix: explicitly setting \texttt{hipStreamDefault} during kernel launch. This parameter was omitted in the original HIP code, causing unintended synchronous behavior. By suggesting this fix, the LLM indirectly resolved a performance bug—demonstrating that \sysname can help detect general inefficiencies even without detailed profiling context. While the fix resembles static correction, the fact that it emerged from performance diagnostics suggests that even limited context can trigger meaningful repairs.

(3) For more complex applications like \sobol and \acc, the LLM suggests optimizations using CUDA-specific syntax (e.g., \newline
\texttt{\_\_restrict\_\_},  \texttt{threadIdx.x}) that are not directly applicable in HIP. As a result, these suggestions either do not compile or fail to yield measurable performance benefits. This reflects a current limitation in the LLM’s training data: it lacks exposure to HIP-specific optimization idioms and AMD-specific hardware behaviors. For example, the LLM may recommend shared memory tuning or warp-level primitives that are implemented differently in HIP or require architecture-specific flags. These gaps do not reflect a flaw in \sysname itself, but rather in the general-purpose model’s hardware-specific knowledge.
%
We believe this limitation can be addressed in future work by integrating more HIP and AMD-specific performance data via retrieval-augmented generation (RAG)~\cite{lewis2020retrieval}, thereby improving the LLM’s architectural grounding while preserving the prompt-driven interface.

\subsection{Explain~\sysname's Reasoning}
\label{sec:reason}
\begin{figure}
    \centering
    \includegraphics[width=\columnwidth]{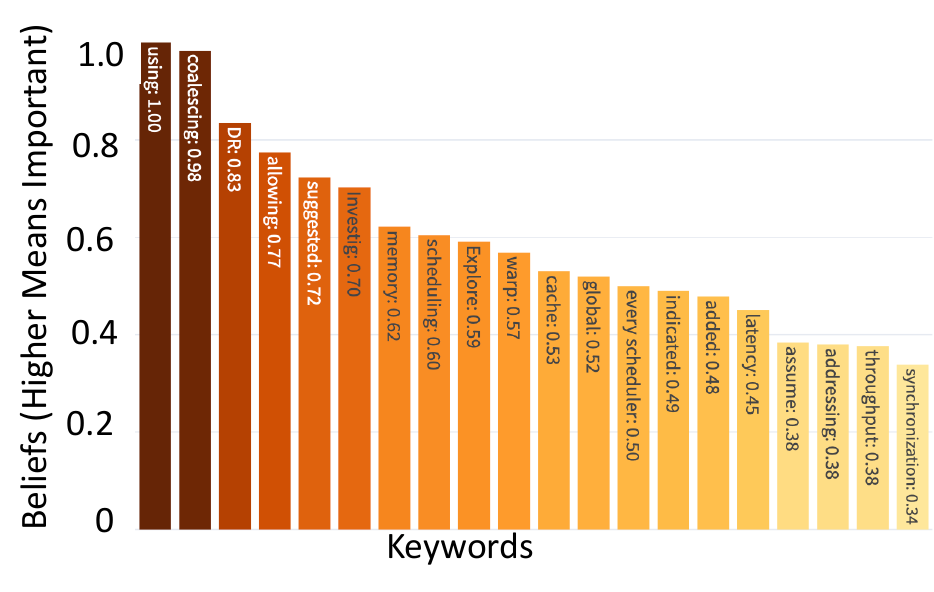}
    \caption{Top 20 keywords and their relative importance. These keywords were deemed important while recommending to use \texttt{\_\_restrict\_\_} for the \acc kernel.}
    \label{fig:reason}
\end{figure}
The objective of this experiment is to explain the reasoning process behind the~\gls{llm}’s code transformation decisions. We analyze belief scores computed from the logprobabilities returned by the~\gls{llm} during inference. These scores reflect which keywords in the prompt had the greatest influence on its final recommendation. To understand why the LLM inserted the \texttt{\_\_restrict\_\_} qualifier in the \acc kernel, we examine its belief distribution in Figure~\ref{fig:reason}. High-scoring keywords include \textit{coalescing}, \textit{pointers}, \textit{memory access}, and \textit{alias}—all terms associated with memory layout and bandwidth utilization. These directly relate to the cache line underutilization problem identified by the Roofline analysis. Interestingly, while \texttt{\_\_restrict\_\_} itself did not appear in the top-ranked tokens, the LLM inferred aliasing as the root cause, linking it to known solutions from the literature. This multi-step inference supports our central claim that by providing structured performance insights to~\gls{llms}, even general-purpose LLMs can reason through the implications and synthesize relevant code optimizations known in the literature.

\subsection{Demonstrate on HPC Proxy Applications}

The objective of this experiment is to evaluate the applicability of \sysname on real-world \gls{hpc} proxy applications that reflect the computational patterns found in national security and scientific workloads. We selected two widely used proxy applications: \texttt{XSBench} and \texttt{sw4lite}.
\texttt{XSBench} models the core loop of Monte Carlo neutron transport, a critical component in nuclear reactor simulation and radiation shielding analysis. \texttt{sw4lite} captures stencil-based wave propagation, a common kernel in seismic inversion and earthquake modeling used for geophysical exploration and hazard assessment. These codes represent two distinct but foundational computation patterns—irregular lookups and structured stencils—found in mission-critical applications ranging from nuclear safety to homeland infrastructure resilience.

For this experiment, we used only two performance insight sources: Roofline analysis and IA. While it is technically possible to apply PC Sampling to these codes by restructuring them into single-file kernels, doing so would require significant manual effort and restructuring—an unrealistic burden in practice. Instead, we chose to investigate whether \sysname can still generate meaningful optimizations using partial diagnostics, reflecting a more practical deployment scenario for large-scale scientific applications.

\textbf{Key observations} 
For \texttt{XSBench}, \sysname increased block size and introduced shared memory usage to reduce global memory pressure and improve data locality. For \texttt{sw4lite}, it unrolled a compute-intensive loop and removed offset computations to reduce register pressure and improve instruction throughput. These optimizations were guided by the combination of Roofline analysis and IA, as PC Sampling could not be applied due to tool limitations on multi-file applications.

Although PC Sampling was unavailable, these results show that \sysname can still operate effectively using partial diagnostics. The improvements—ranging from 0.00\% to 11.58\% for \texttt{XSBench} and 13.15\% to 14.56\% for \texttt{sw4lite}—suggest that combining architectural modeling with low-level interaction profiling is sufficient to trigger meaningful transformations in large, real-world codebases.

\subsection{Discussions}
\textbf{Overhead} The overhead of recommending code transformations is a few seconds for all experiments. However, the data collection time can be extensive for PC Sampling source. 
Since these tools and methods are established in the community~\cite{RooflineScalingTrajectories,lee2020roofline,williams2009roofline, amd_roofline_micro2022,lawson2015experimentationprocedureoffloadedminiapps,zhou2021gpa,zhou2020tools,islam2016a,dashing,islam2019toward,zaeed2024characterize,islam2025data,banerjee2016cmtbone}, we assume that users interested in optimizing their most time consuming kernel's performance will start with a subset of these tools. Perhaps future research can synthesize these profiles for different regimes from a few runs instead of repetitively running the code. 

As discussed in Section~\ref{sec:defaultconfig}, \sysname's explanations made it possible to identify an unsafe transformation and swap it with a safer one. 
With \sysname, users can easily validate, revert, or substitute code transformations without needing to design them from scratch. 

\textbf{One size does not fit all}
Some GPU kernels (e.g., \texttt{accuracy}) show consistent bottlenecks across configurations, while others may vary based on input regimes. In this work, to keep validation tractable, we use \sysname to optimize for one configuration per application. Despite this simplification, we observe that these optimizations generalize across configurations. 
In the future, we will building on \sysname to investigate novel approaches for clustering input regimes based on bottlenecks and generating tailored optimizations per group. 
Without \sysname, mapping from bottlenecks to code edits would remain a manual trial-and-error process. 

\textbf{More data is not always better: prompt tuning matters}
Our experiments show that combining multiple analysis sources helps. However, longer prompts often cause \glspl{llm} to focus on less relevant bottlenecks. This issue stems from prompt token limits and is compounded by the fact that hardware counters often lack clear, descriptive explanations. Even so, \sysname consistently improves performance across applications. These findings highlight that success depends not just on the data, but on 
the quality of the insights extracted by \sysname.
In the future, we will explore adaptive prompting strategies that identify the most relevant diagnostics across sources. 

\section{Related Work}
\label{sec:related-work}
Performance analysis methodologies such as Roofline modeling~\cite{williams2009roofline,RooflineScalingTrajectories,lee2020roofline,amd_roofline_micro2022,Ilic14CRM}, PC Sampling~\cite{zhou2021gpa,chabbi2013effective,zhou2020tools,zhou2021measurement} and hardware performance counters~\cite{uhsadel2008exploiting, islam2019dashing,islam2016a} help identify architectural inefficiencies (e.g., memory bottlenecks, TLB misses) but require manual interpretation and expert-guided code tuning. Similarly, PC Sampling tools such as Nsight~\cite{nvidia2023nsight,zhou2021gpa} localize stalls to source lines but leave the optimization step to the user. While these approaches are effective at diagnosis, they do not automate the path to resolution. \sysname provides an extendable framework that complements this ecosystem by structurally translating these diagnostics into insights that can guide \gls{llms} to generate actionable code changes, thus closing the loop between profiling and code transformation.

General-purpose models such as Codex~\cite{chen2021evaluatinglargelanguagemodels}, StarCoder~\cite{li2023starcoder}, and CodeLlama~\cite{roziere2023code} are trained for code completion and refactoring but lack awareness of runtime performance bottlenecks. Similarly, Star-Agents~\cite{zhou2024star} learn from version histories but do not incorporate profiling data. As a result, these models tend to propose generic transformations that are disconnected from actual hardware constraints. \sysname addresses this gap by capitalizing on the reasoning capability of the \gls{llms} to connect optimization strategies to bottlenecks commonly found in the literature.

Autotuners such as OpenTuner~\cite{ansel2014opentuner} and learning-based tools such as OptFormer~\cite{luo2023optformer} automate code-space exploration via heuristic search or policy learning. However, they rely on manual knob definition or optimize for proxy objectives such as loop unrolling depth or binary size. In contrast, \sysname derives optimization intent directly from multi-source performance diagnostics and maps them to meaningful code transformations using off-the-shelf LLMs. To our knowledge, \sysname is the first framework to integrate diverse analytic sources into an automated code optimization engine.

\section{Conclusions}
\label{sec:conclusions}
Despite the availability of many \gls{hpc} performance analysis tools, the final step of translating diagnostic outputs into actionable optimizations remains manual and expert-driven. We introduce \sysname to bridge this gap by transforming multi-source performance insights into structured prompts for \gls{llms}. These natural language prompts enable \gls{llms} to reason about which code transformation to apply based on their knowledge of the literature. Across $\sim$1640 experiments, \sysname achieved performance improvements ranging from 1.76\% to 87.6\%, with only 1 out of 1640 experiments failing to pass the validation tests, greatly highlighting that \sysname is able to use the \gls{llms} as a reasoning engine. 
By combining performance analytics with LLM-guided transformation, \sysname takes a significant step toward democratizing expert-level GPU performance engineering.
In future work, we aim to extend \sysname into a fully integrated multi-agentic system that automates the invocation and execution of performance analysis tools, completing the loop from profiling to synthesis.

\bibliographystyle{ACM-Reference-Format}
\bibliography{tzi}


\end{document}